# A DDS-BASED SCALABLE AND RECONFIGURABLE FRAMEWORK FOR CYBER-PHYSICAL SYSTEMS

Ismael Etxeberria-Agiriano[1], Isidro Calvo[2], Liliana Montero[1] and Ivan Alonso[1]

[1] Dept. of Computer Languages and Systems, EUI Vitoria-Gasteiz, UPV/EHU Spain
ismael.etxeberria@ehu.es

[2] Dept. of Automatic Control and Systems Engineering, EUI Vitoria-Gasteiz, UPV/EHU Spain
isidro.calvo@ehu.es

## ABSTRACT

*Cyber-Physical Systems (CPSs) involve the interconnection of heterogeneous computing devices which are closely integrated with the physical processes under control. Often, these systems are resource-constrained and require specific features such as the ability to adapt in a timeliness and efficient fashion to dynamic environments. Also, they must support fault tolerance and avoid single points of failure. This paper describes a scalable framework for CPSs based on the OMG DDS standard. The proposed solution allows reconfiguring this kind of systems at run-time and managing efficiently their resources.*

## 1. INTRODUCTION

Cyber-Physical Systems (CPSs) are embedded computers and networks which interact with physical processes integrating computation, communications and the dynamics of physical processes. Applications of CPSs include automotive and traffic control, healthcare, factory automation, infrastructure control (e.g. electricity, water and communications), distributed robotics and process control [1]. Shi et al. [2] describe the main features of CPSs as well as showing three classical application domains. CPSs are considered a next step forward in control and computing co-design. Unfortunately, there exists a surprisingly small amount of theory about how to design computer-based control systems [3]. Wu et al. [4] review some research activities in the field of wireless sensor networks as applied to CPSs.

However, CPSs face many challenging problems such as fault tolerance or reconfiguration. In these systems individual elements may fail but the whole distributed system must be reliable and capable to perform correctly even in degraded reconfigured modes. Maintenance must be easy and affordable.

CPSs use intensively communications, which in most cases are based over IP protocols [5]. However, as the complexity and size of CPSs grows, the use of distribution middleware solutions is desirable to help developers to manage the use of communications among the distributed nodes that may be executed over heterogeneous hardware platforms and operating systems. Some examples of well established distribution middleware technologies that allow the integration of the devices involved in CPSs are CORBA [6, 7], OPC [8] or Web Services [9]. In particular, DPWS (Devices Profile for Web Services) [10], defines a specification of Web Services aimed at devices with low resources that allows secure Web service messaging, discovery, description and eventing on resource-constrained devices.

More recently, the Object Management Group (OMG) released DDS (Data Distribution Service for Real-Time Systems) [11], which is a data centric middleware specification which follows the publisher/subscriber paradigm. DDS is a unique standard middleware specification that allows combinations of different programming languages (mainly C, C++ and Java), operating systems (such as Windows and Unix/Linux derivatives) and hardware platforms. The DDS specification is aimed at providing a high-performance publisher/subscriber middleware layer which is highly configurable through a rich set of QoS (Quality of Service) parameters. This is particularly interesting in contexts where communication services with different requirements coexist. Table 1 provides a list of the most relevant DDS QoS policies.

Table 1. DDS Quality of Service Policies.

| | | | |
|---|---|---|---|
| Deadline | History | Partition | Time-Based Filter |
| Destination Order | Latency Budget | Presentation | Topic Data |
| Durability | Lifespan | Reader Data Lifecycle | Transport Priority |
| Durability Service | Liveliness | Reliability | User Data |
| Entity Factory | Ownership | Resource Limits | Writer Data Lifecycle |
| Group Data | Ownership Strength | | |

DDS is a valuable technology for CPSs since it provides a platform-independent middleware layer for Data Centric Publish/Subscribe many-to-many communications. Actually, several authors such as [12, 13, 14, 15] propose solutions on top of this specification. DDS has been designed so that assuming that the underlying network infrastructure is deterministic, the middleware layer does not introduce indeterminism. Furthermore, DDS allows the management of many aspects of the communication behaviour to meet application requirements. As opposed to message-oriented approaches, DDS does not exchange data in the form of messages, but data is formally defined in a platform independent way. This approach allows the automatic generation of source code for specific target platforms. In addition, since DDS follows the publisher/subscriber paradigm, it allows decoupling producers and consumers of information. As a consequence, DDS may be seen as a kind of virtual bus software which abstracts the underlying network infrastructure.

This paper describes a reconfigurable framework that uses DDS as distribution middleware and exploits some of its features. Namely, it benefits from its efficiency and the rich set of QoS management properties. It is important to remark that the proposed framework must coexist with the distributed tasks that define the functionality of the application. More specifically, it provides an additional layer on top of DDS with mechanisms for managing fault-tolerance, reconfiguration to adapt to changes both in the functionality and in the environmental conditions, and resource management of the distribution application nodes. Finally, this framework improves the scalability of the system.

The reminder of this paper is structured as follows. Section 2 describes some preliminary concepts related to the proposed framework and its requirements in CPSs. These are the basis for the construction of a family of virtual topologies as detailed in Section 3. These topologies are used in Section 4 to maintain global state information using the DDS middleware. The evaluation of this framework is described in Section 5 simulating Poisson distribution task activations. Finally, Section 6 draws conclusions and proposes future evolution threads of this research.

## 2. PRELIMINARY CONCEPTS

This Section introduces some preliminary concepts that will be used hereafter. It must be noted that the framework cohabitates with the distributed applications. So the messages required by the framework will share the same network infrastructure as the messages exchanged by the distributed application. Consequently, DDS can be considered as a software virtual bus for the whole CPS.

### 2.1. Resource Availability Modelling

Reconfiguring a CPS requires some criteria. Typically, one reconfiguration criterion is based on the resource availability in order to determine which nodes are best suited to carry out a task. Therefore, it is necessary to model the available resources Examples of potential resources are CPU, memory or battery [16].

A clairvoyant ideal system knows at each instant the resource capacity of every node. Brought to practice, this approach introduces some drawbacks, especially as the complexity of the system increases. Also, it may be difficult to maintain in dynamic environments.

For the sake of describing the resource availability the term **status** is used here. It is assumed that the underlying operating systems provide mechanisms for measuring the values of these resources by the local applications at run-time.

In order to refresh the resource status information two approaches can be followed: (1) event triggered, that is, updates are pushed by significant resource status changes or (2) time triggered, that is resource status is periodically updated.

### 2.2. Fault Tolerance and Reconfiguration

Some CPSs require fault-tolerance mechanism, whereas in other cases it is a desired design feature. A basic scenario can be a device running out of batteries. If the device is capable of detecting that a battery threshold is surpassed, it may signal it to its neighbors so one of them replaces it without experiencing service disruption. Otherwise, a periodic heartbeat may be used to check whether all devices are alive.

## 2.3. Process migration and data distribution

Dynamic CPSs may require the migration of code, data or running processes from one node to another. In addition, these mechanisms may be used for other features such as installing or updating new software versions at run-time. Some authors present solutions for this problem, some of them on top of DDS. An example may be found in Park et al. [13].

Also, appropriate mechanisms must be set to distribute relevant data to the processing unit. A data-centric middleware like DDS is very adequate for this purpose, since it provides QoS control mechanisms. As a matter of example, DDS allows reassigning ownership on data readers and writers which, combined with the persistence QoS, may be employed as useful and efficient strategies to redirect data from a faulty device to its replacer.

## 2.4. Reconfiguration Candidate Selection

There are multiple situations that may trigger reconfiguration. Sometimes, the failure or the replacement of a node will require a quick solution to avoid that some running functionalities remain unattended. More frequently, resource usage optimization or environmental changes can motivate reconfiguration.

In a reconfiguration stage it may be necessary to decide which node is the best candidate to take over to carry out a task. This process is known as candidate selection.

There are two main candidate selection types of approaches: some are (1) sender initiated, when the nodes requiring a service initiate the reconfiguration process. However, this strategy has no reference of current system status. Additionally, together with a delay, it also introduces an extra overhead when the overall system is highly loaded, which is not desirable. Other approaches are (2) receiver initiated. In this case, nodes with spare resources offer services that are used only if necessary. One interesting option is to establish a virtual communication topology to provide the identity of nodes capable of taking over some tasks.

## 3. STATUS INFORMATION EXCHANGE STRUCTURE

This section introduces a family of information exchange structures called **Reliable Friend** (RF). This name comes intuitively from the idea that, in its simplest version, each node will try to keep the identity of one node with spare resources to replace it. The reliable candidate will be based on the resource availability information. The advantage of limiting this information to one or just a few nodes is that information exchange is reduced, producing highly scalable systems. In CPSs with hundreds or even thousands of participant nodes, nodes are given a local view of an alternative to solve failures or temporal high workload situations.

It is based on a simple structure. Some alternatives are also described to illustrate the potential of these structures, which never rely on a centralized single point of failure entity.

Simulations of the behaviour of this algorithm show that it is an effective approach when compared to non-cooperative, clairvoyant and random variants [17].

## 3.1. Basic Reliable Friend

The relationship among nodes can be established based on the bootstrap hazard or on some communication convenience, among other possibilities. This relationship will determine the reconfiguration target preference for each node, resulting in a virtual topology called the **Official Structure** (OSt).

A directed cyclic graph has been chosen as the basic OSt. Fig 1.a depicts the OSt of a system with 10 participating nodes. Under resource availability node $n_0$ will rely on node $n_1$; node $n_1$ will rely on $n_2$, and so forth. With this naming convention it can be seen that for m nodes, node $n_{m-1}$ is linked to $n_0$. Under this relationship, two nodes $n_i$ and $n_{i+1}$ are called **Official Sender** (OSe) and **Official Receiver** (ORe) respectively.

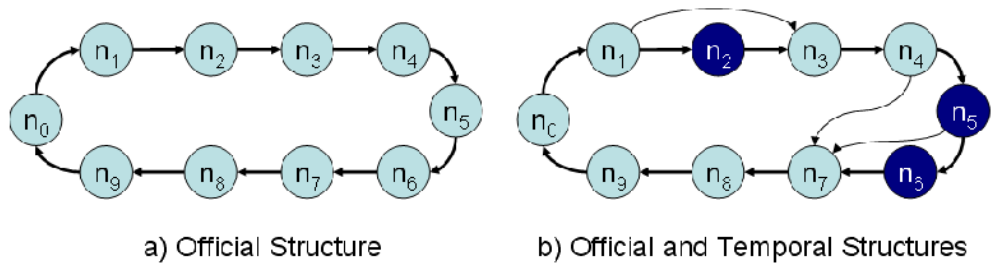

Figure 1. Example of Basic Reliable Friend Structures with 10 nodes.

The willingness of nodes to receive or send workload from/to other nodes will depend on an application dependent resource availability status, introduced in the previous Section. In the basic case two possible states are considered: **Available**, or able to receive processes, and **Unavailable**, or ready to send processes. From a node's local point of view time or event triggered evaluation is required in order to detect transitions from Available to Unavailable and vice versa.

A status information exchange mechanism can be used to establish a dynamic structure called the **Temporal Structure** (TSt). Figure 1.b depicts the TSt of a system with 10 participating nodes where some of them, namely $n_2$, $n_5$ and $n_6$, are represented as Unavailable in dark colour. As a consequence of this dynamic situation, nodes $n_1$, $n_4$ and $n_5$, cannot rely on them and they need a **Temporal Receiver** (TRe). It has been chosen that the TRe for a node is the first Available node found following the OSt.

## 3.2. Bidirectional Reliable Friend

Sometimes having just one candidate is not enough to ensure high reliability, especially when states change rapidly or the information is not regularly updated. On top of that, as a result of reconfiguration, Unavailable nodes may accumulate resulting more hazardous.

An alternative to the basic RF structure can be the **Bidirectional RF** (BRF) structure. The OSt, as depicted in Figure 2.a, provides the information of two official candidates for each node. The TSt (Figure 2.b) is maintained in such a way that the identity of two Available nodes is kept, whenever there is any present in the system.

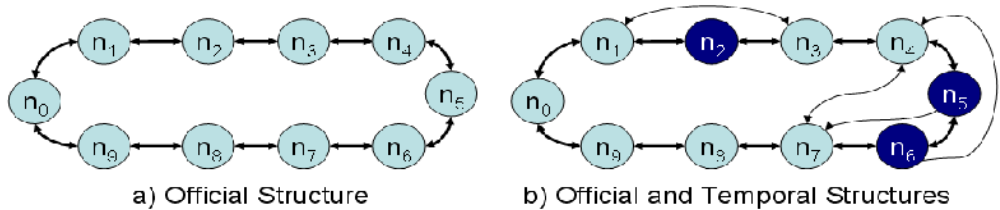

Figure 2. Bidirectional Reliable Friend Structure with 10 nodes.

### 3.3. Simplified Bidirectional Reliable Friend

The BRF provides the identity of two participant candidates for each nodes but maintaining the structure can be more time consuming. The **Simplified BRF** (SBRF) reduces this information update in the case of Unavailable nodes, so that for Unavailable nodes the identity of only one candidate is kept. In Figure 3 it has been chosen that the TRe for a node to be the closest node (in number of hops) in both directions, and the clockwise direction to break ties. The intention of this alternative is that allow Unavailable zones expand in both directions.

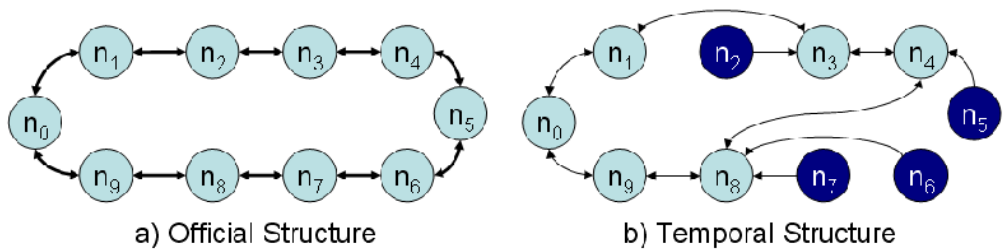

Figure 3. Simplified Bidirectional Reliable Friend Structures with 10 nodes.

### 3.4. Other Reliable Friend Variants

Other variants of RF structures have also been considered which are more relevant for load balancing, namely: Multi-Level RF (MLRF) and Multi-State RF (MSRF). Combinations of RF structures can also be used in the form of Multi-Purpose RF and Multi RF. As they do not add much to the main purpose of this paper they have been omitted. A more detailed description of these topologies can be found in [18].

## 4. FRAMEWORK IMPLEMENTATION DETAILS OVER DDS

Previous Section presented a family of reconfiguration topologies called Reliable Friend (RF). This Section discusses the RF framework implementation details on top of DDS. In particular, it is discussed the definition of the topics to keep the topologies and the selection of the DDS QoS parameters.

## 4.1. Reliable Friend Official Structure Construction

Figure 4 illustrates the construction of the Official Structure (OSt) under different scenarios. First case in Figure 4.a occurs when a node $n_0$ is alone in the system, resulting in a single node structure. Under that situation, when another node $n_1$ arrives, as in Figure 4.b, it will become a two node structure. Figure 4.c describes the situation where a two node structure exists and a new node, $n_2$, arrives. It illustrates how it is inserted at the end of the structure following the arrival sequence. This process continues as nodes incorporate.

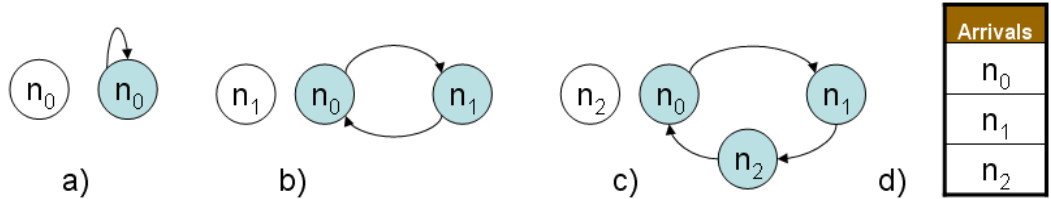

Figure 4. Insertion of a node a) $n_0$ with no others; b) $n_1$ with another; c) $n_2$ with other nodes; d) the Arrivals topic keeps a history of the identity in arrival order of m participants.

In order to apply this process with DDS, a topic called **Arrivals** (Fig 4.d) is used with a history QoS value set to a value *m* high enough. An incoming node, $n_i$, will publish its identity on this topic and it will immediately after subscribe to it. On the first case, shown in Figure 4.a, node $n_0$ will only find its own identity and it will form the single-node structure. In other cases nodes $n_1$, $n_2$, … will find on this topic the identity of the node inserted just before themselves, i.e. $n_0$, $n_1$, …

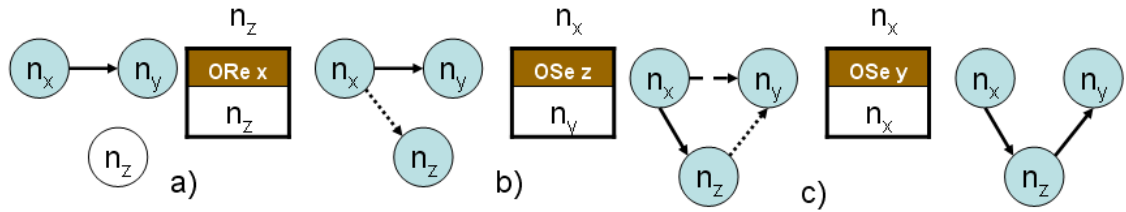

Figure 5. Insertion of a node a) $n_z$ notifies $n_x$; b) $n_x$ notifies $n_z$; c) $n_x$ notifies $n_y$;

To actually update this structure the action sequence shown in Figure 5 will be followed. In Figure 5.a node $n_z$ wants to insert itself between $n_x$ and $n_y$ only knowing the identity of $n_x$. $n_z$ publishes its identity on a topic where $n_x$ waits subscribed for new **ORe**-s. Then $n_x$ will update this information and publish two topics: one informing on the new **OSe** for $n_z$ (Figure 5.b) and another (Figure 5.c) informing on the new **OSe** for $n_y$.

## 4.2. Reliable Friend Temporal Structure Construction

Independently from the number of participating nodes the vision of the system for any of the nodes is limited to itself (Me), its Official Sender (OSe), its Official Receiver (ORe), and sometimes its Temporal Receiver (TRe). This is illustrated in Figure 6 under different scenarios where any of the nodes, Me, can be the reference. For any node Me we may find that: Figure 6.a its ORe is Available, so that it can be trusted; no state information on the OSe is present; Figure 6.b its ORe is Unavailable; the identity of the next Available is present, with no reference on how many Unavailable nodes are in between; Figure 6.c all nodes but itself, Me, are Unavailable; when this node becomes Unavailable the situation in Figure 6.d may arise, i.e. all nodes become Unavailable; this situation can be detected as in Figure 6.c it detects that Me was the only Available node.

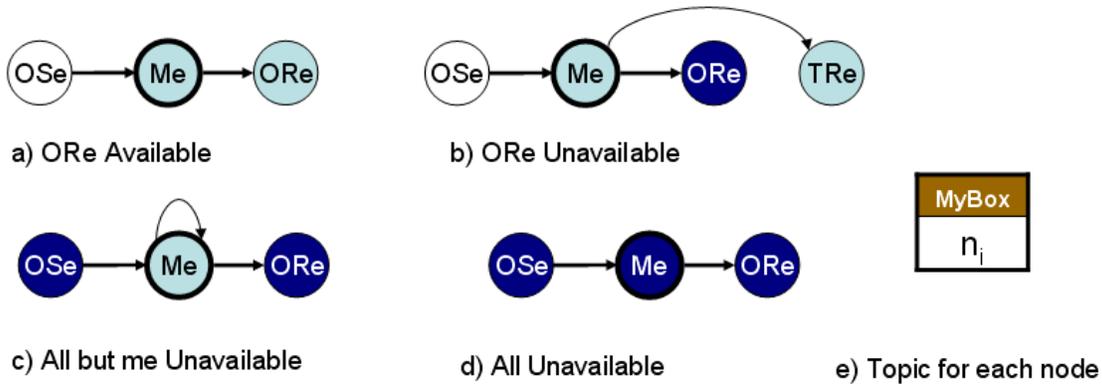

Figure 6. Vision of the system state from the point of view of a node (Me)

In order to maintain this information consistently all nodes keep the following information:

- Me: my identity, identity of the node concerned
- ORe: identify of the Official Receiver (possible Me)
- TRe: identity of the Temporal Receiver (possibly Me or Null)
- OSe: identity of the Official Sender
- State: my availability state, Available or Unavailable

Nodes may publish and subscribe to DDS topics as necessary. All nodes $n_i$ **publish** a topic MyBox$_i$ with the identity of the first Available node in the sub-graph starting from itself.

All nodes always **subscribe** to the topic of the following node, the ORe. Additionally, whenever the ORe of a node is Unavailable, nodes also **subscribe** to the topics published by the TRe as it may provide updating information of the identity of another node further in the directed graph.

Three possible events may occur that make this structure change: a) the transition of a node from Available to Unavailable, b) the transition of a node from Unavailable to Available and c) the reception of a topic update, which may cause the propagation of Msg, the identity of an Available node or Null when such a node is not present in the whole system. This latter case allows distinguishing two cases: one coming from the ORe and another coming from the TRe, although both could be treated in a uniform manner.

**4.2.1. State Change from Available to Unavailable**

When a node becomes Unavailable and it was the last Available node in the system, then no Available node remains in the system. This condition can be checked just by verifying if the TRe corresponds with its own identity. The identity of the previous TRe or Null if it was itself is published on MyBox and this information is received by all nodes subscribed to it, being its OSe and many possible temporal senders. This action can be represented algorithmically:

> **If** TRe = Me **Then**
>   TRe ← **Null**
> Publish (MyBox, TRe)

**4.2.2. State Change from Unavailable to Available**

When a node becomes Available, it just needs to inform its OSe, which will need to restore its TRe to its ORe. This information will be propagated backwards in the Official Structure as necessary. DDS allows a special case when all nodes become Unavailable, which is subscribing all to a special **OneBack** topic, waiting for the first node to become Available. Instead of relying on a one-by-one propagation this would profit the DDS distribution mechanism which is expected to perform efficiently. The first node becoming Available will publish on this topic. This new case can be represented algorithmically:

```
If TRe = Null Then
    Publish (OneBack, Me)
    TRe ← Me
Publish (MyBox, Me)
```

**4.2.3. Topic Update on ORe, with a New Msg ←TRe**

This update means either that the ORe has become Available and we get its identity or that the identity of the TRe has changed. This may result from a status modification of our ORe or some node found following the cyclic graph. If the node was subscribed to a TRe it must cancel this subscription; if the identity received is not that of its ORe the corresponding subscription must be issued. Finally if the node is Unavailable this information needs to be propagated backwards. An exception must be considered when its identity has been propagated all around the directed graph. In that particular case propagation is stopped. This can be represented algorithmically:

```
If TRe   Null Then
  CancelSubscription(TRe)
If Msg = ORe Then
  TRe ← Null
Else
  Re ← Msg
Subscribe (TRe)
If (State = "Unavailable" And Msg   Me) Then
  Publish (MyBox, Msg)
```

**4.2.4. Topic Update on TRe, with a New Msg ←TRe**

This update is similar to the previous one but always means that the current TRe of the node receiving Msg has become Unavailable. The previous subscription is cancelled and a new subscription is issued if Msg carries the identity of an Available node. No propagation is required as all Unavailable nodes before the updating node are subscribed to its topic. This is represented algorithmically:

```
CancelSuscription(TRe)
TRe ← Msg
If TRe   Null Then
   Subscribe (TRe)
```

## 4.3. Quality of Service Configuration

The construction and maintenance of the RF structure require 5 communication topics:

1. **Arrivals**: used by arriving nodes to determine their ORe
2. **ORe**: used by arriving nodes to notify a change in the OSt
3. **OSe**: used by a notified node to rearrange the OSt
4. **MyBox**: used by nodes to publish the TRe
5. **OneBack**: used by a node becoming Available when no one is known to be Available

The main DDS QoS configuration parameters used for these services are summarized in Table 2. All these services will be combined with the application dependent services.

Table 2. Mapping of DDS Quality of Service Policies on the RF services

| | Topics | | | |
|---|---|---|---|---|
| | **Arrivals** | **ORe & OSe** | **MyBox** | **OneBack** |
| **Distribution** | One to many | One to one | One to many | One to many |
| **Destination order** | Source | Source | Source | Source |
| **Durability** | Persistent | Volatile | Volatile | Volatile |
| **History** | Keep N | Keep last | Keep last | Keep last |
| **Latency Budget** | Life proof | Life proof | Life proof | Life proof |
| **Lifespan** | Long | Short | Short | Short |
| **Liveliness** | Automatic | Automatic | Automatic | Automatic |
| **Reliability** | Reliable | Reliable | Reliable | Reliable |
| **Transport Priority** | Highest | Highest | Highest | Highest |

## 5. FRAMEWORK EVALUATION

For the evaluation of the framework construction method presented in previous section several simulations were implemented with some variants [19]. The test scenario consisted of four real nodes running the simulation program developed using OpenSplice [20] but could be used with an arbitrary number of nodes.

The Basic Reliable Friend (RF) Official Structure (OSt) was constructed according to the proposed mechanisms and some tests were carried out to verify that the circular ring was correctly obtained. It is noted that this construction schema only works in absence of failures. This assumption is considered acceptable for the sake of testing the basic construction method and the dynamics behaviour of the temporal structure. An improved design should be complemented with failure detection and reconfiguration mechanisms.

The Basic RF Temporal Structure (TSt) construction test implies the simulation of task activations. Following the fundamentals of the Basic RF TSt construction two states are considered, (1) Available, or able to assume more workload than the one already assigned; and (2) Unavailable, or unable to assume more workload.

Poisson distribution (1) task activation was simulated where k is the number of occurrences of the phenomenon in a given interval and $\lambda$ is a positive parameter that represents the number of times the event occurs in that interval.

$$f(k; \lambda) = \frac{e^{-\lambda} \lambda^k}{k!},$$

(1)

This probabilistic variable was utilised with a threshold determining the availability for each node on next interval. Nodes simulated these random state changes and their corresponding information updates.

The proposed mechanisms worked correctly on the different evaluation tests carried out on the final design.

## 6. CONCLUSIONS AND FUTURE WORK

This work has presented a scalable and reconfigurable framework on top of DDS for CPSs. DDS has been selected since it is unique and high performance middleware specification that provides fine grained mechanisms to manage the QoS properties of communications. Also, it solves the problem of heterogeneity since it can coexist with different programming languages, hardware platforms and operating systems. The proposed framework, which is relatively simple, is capable of adapting in a timeliness and efficient way to dynamic environments. Also, it supports fault-tolerance and avoids single points of failure. Finally, it eases the management of the resources of the CPSs and allows the reconfiguration of the system at run-time in response to node failures, changes in the environment, modifications in the functionality of the system or software updates.

The proposed framework was evaluated using the commercial OpenSplice DDS product showing an adequate behaviour in the initial construction of the RF official structure and the dynamic maintenance of the RF temporal structure.

The envisioned future work includes measuring the utilized bandwidth and incurred delays on distributed applications executed on embedded platforms. Also, the benefits of migrating code among computers and using redundant services will be analyzed. Finally, the capacity of detecting failures and reconfiguring the official and temporal structures are also point of future interest in this research, together with the application task data.

## ACKNOWLEDGEMENTS

This work was supported in part by the Basque Government (Saiotek) under project S-PE11UN061 and the University of the Basque Country through the grant EHU11/35.

## AUTHORS

**Ismael Etxeberria-Agiriano** obtained a degree in Electronic Engineering and a degree in Computer Science from the Mondragon Unibertsitatea (MU). In 1994 he obtained his PhD from Staffordshire University (SU). He worked as a consultant for the Crédit Lyonnais in Paris (France) and continued working on financial applications for another three years before returning to education. He has been a lecturer at both SU and MU universities. Since year 2005 he is contracted as lecturer at the Computer Languages and Systems Department of the University College of Engineering of Vitoria (UPV/EHU). His main research interests include Distributed Systems, Computer Security and Applied Ethics. 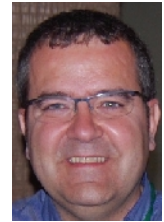

**Isidro Calvo** obtained a degree in Physic Science in 1993 from the University of the Bas que Country (UPV/EHU). In 1994 he completed a Master degree in Electronics and Automatic Control, at the same University. He worked as a software engineer for several companies in Spain and UK between 1995 and 1999. In 1999 he joined UPV/EHU as a researcher obtaining his PhD in 2004. Since year 2000 to 2007 he was contracted as Lecturer at the Automatic Control Department of the Engineering School of Bilbao. Nowadays, he is with the University College of Engineering of Vitoria (UPV/EHU) as Senior Lecturer. He has collaborated in several research projects funded by National & European projects co-authoring several technical papers in international journals and conference proceedings. His main research interests include Middleware, Embedded systems and Remote Learning. 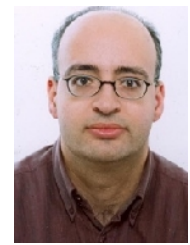

**Liliana Montero-Fernandez** obtained a degree in Computing Engineering in 2011 from the University of the Basque Country (UPV/EHU).

**Ivan Alonso-del-Val** obtained a degree in Computing Engineering in 2011 from the University of the Basque Country (UPV/EHU).